\documentclass[aip,amsmath,amssymb,reprint]{revtex4-1}
\usepackage{graphicx}
\usepackage{dcolumn}
\usepackage{bm}
\usepackage[utf8]{inputenc}
\usepackage[T1]{fontenc}
\usepackage{mathptmx}
\usepackage{subcaption}
\usepackage{tikz}
\usetikzlibrary{shapes,arrows}
\usepackage{booktabs}
\begin{document}

% Use the \preprint command to place your local institutional report number 
% on the title page in preprint mode.
% Multiple \preprint commands are allowed.
%\preprint{}

\title{Embedded model discrepancy: A case study of Zika modeling} %Title of paper

% repeat the \author .. \affiliation  etc. as needed
% \email, \thanks, \homepage, \altaffiliation all apply to the current author.
% Explanatory text should go in the []'s, 
% actual e-mail address or url should go in the {}'s for \email and \homepage.
% Please use the appropriate macro for the type of information

% \affiliation command applies to all authors since the last \affiliation command. 
% The \affiliation command should follow the other information.

\author{Rebecca E. Morrison}
\email[]{rebeccam@colorado.edu}
\homepage[]{rebeccaem.github.io}
%\thanks{}
%\altaffiliation{}
\affiliation{Department of Computer Science, University of Colorado Boulder, Boulder, CO, USA 80309}

\author{Americo Cunha Jr}
\email[]{americo@ime.uerj.br}
\homepage[]{www.americocunha.org}
%\thanks{}
%\altaffiliation{}
\affiliation{Department of Applied Mathematics, Rio de Janeiro State University (UERJ), Rio de Janeiro, Brazil 20550}

% Collaboration name, if desired (requires use of superscriptaddress option in \documentclass). 
% \noaffiliation is required (may also be used with the \author command).
%\collaboration{}
%\noaffiliation

\date{\today}

\begin{abstract} Mathematical models of epidemiological systems enable investigation of and
    predictions about potential disease outbreaks. However, commonly used models are often highly
    simplified representations of incredibly complex systems. Because of these simplifications, the
    model output, of say new cases of a disease over time, or when an epidemic will occur, may be
    inconsistent with available data. In this case,  we must improve the model,
    especially if we plan to make decisions based on it that could affect human health and safety, but
    direct improvements are often beyond our reach. In this work, we explore this problem through a
    case study of the Zika outbreak in Brazil in 2016.  We propose an embedded discrepancy
    operator---a modification to the model equations that requires modest information about the
    system and is calibrated by all relevant data. We show that the new enriched model demonstrates
greatly increased consistency with real data. Moreover, the method is general enough to easily apply
to many other mathematical models in epidemiology.  \end{abstract}

% insert suggested PACS numbers in braces on next line

\maketitle %\maketitle must follow title, authors, abstract and \pacs

% Body of paper goes here. Use proper sectioning commands.  
%References should be done using the \cite, \ref, and \label commands

\textbf{Potential epidemics of communicable diseases  are a major health concern of the modern
world, especially as city density, air and water pollution, and worldwide travel steadily increase.
A stark example of this is the global Coronavirus outbreak, already responsible for more than 2000
deaths around the world at time of this article's submission, and more than 13,000 at time of
revision, approximately one month later. When faced with a potential
outbreak, decision-makers such as health officials and medical professionals rely on mathematical
models to aid their decision-making processes. But oftentimes these models are not consistent with
the dynamical system they are designed to represent.  The discrepancy between the output of a model
and the real system is then a serious impediment, as it may decrease our confidence in the model, or
even invalidate it entirely, so that it can no longer be used to aid in decision-making.  When
such a discrepancy is observed, we, as modelers, must either improve the model or somehow account
for the discrepancy itself. While a direct model improvement is usually the most desirable solution,
how to do so may be infeasible because of computational reasons, time constraints, or lack of domain
knowledge. This paper provides a systematic method to instead account for the discrepancy itself,
explored via a case study of the Brazilian Zika epidemic of 2016.  The method is not a correction of
the model output to data, but rather a modification of the model equations themselves by a so-called
embedded discrepancy operator. The operator is designed with three critical properties in mind:
interpretability, domain-consistency, and robustness. We show that including the embedded
discrepancy operator greatly increases the fidelity of the model, so that model output and real data
are now in fact consistent.}

\section{Introduction}\label{sec:int} Mathematical models of scientific systems necessarily include
simplifications about the actual system they aim to represent. In some cases, these simplifications
do not preclude the use of the model to understand, investigate, and make decisions and predictions about the
system. The quintessential example of this comes from the domain of classical mechanics: Newtonian
mechanics ignores quantum and relativistic effects but, over a wide domain of masses and energies,
provides a completely adequate model to describe the motion of macroscopic objects. Outside this
domain, however, quantum or relativistic effects are no longer negligible, and Newtonian mechanics
fails.

Just as classical mechanics is insufficient to describe a quantum system, the simplifications of a
modern mathematical model may yield a discrepancy between the model and the system at hand too great
to be ignored.  This discrepancy is revealed during \emph{model validation}, a process by which we
check that the mathematical model is a reliable representation of reality. Without accounting for
the discrepancy, one cannot trust the model output, much less use it to make predictions or
decisions. In this case, there are two immediate options: (1) Improve the model directly, i.e., from
first principles or by including additional information; and (2) Represent the model discrepancy
itself. While option 1 is usually desirable, it may not be feasible due to computational
constraints, or because we in fact lack the knowledge to directly improve the model.  Then we are
left with option 2---represent the model discrepancy. 

A common approach to account for model discrepancy is through a \emph{response discrepancy function}
\cite{kennedy2001bayesian}, also called a bias function. A response discrepancy function corrects
model output (or response) to data. Typically, an additive function on the model output is
calibrated to data, either point-wise or with a parametric form. An advantage of this approach is
that it can be implemented even if the model is a black box, that is, one only needs access to model
output, not the model itself. There are also disadvantages. In essence, a response
discrepancy function builds a better interpolation to a single dataset, over the range of
usable data. Thus, this approach provides no basis for extrapolation, to, for example, make a
prediction about the probability of an epidemic next year. Furthermore, the action of this bias
function is not interpretable, as it lies outside the model equations.

Instead, in this paper, we show how to modify equations directly to account for the model error with
an \emph{embedded discrepancy operator}. The advantages of this approach are threefold:
\begin{enumerate}
    \item \textbf{Interpretability:} As the embedded operator appears within the model equations,
        and acts on state variables, the action of this operator is interpretable.
    \item \textbf{(Domain-)Consistency:} Information or constraints about the system can be
        incorporated into the discrepancy operator.
    \item \textbf{Robustness:} Discrepancy parameters can, and should, be calibrated over all available
        data. This can include data from multiple scenarios or initial conditions. 
\end{enumerate}
These three properties---interpretability, consistency, and robustness---are designed to allow for
decisions or extrapolative predictions. The inclusion of the embedded discrepancy operator into the
original, or reduced model, yields an \emph{enriched model}. In essence, the enriched model takes
advantage of both mechanistic and statistical modeling: it retains the reduced mechanistic model,
and incorporates a general, statistically calibrated discrepancy model. Of course, this intrusive
approach highly depends on the context.  Here, we investigate the value of an embedded discrepancy
operator in the context of epidemiology modeling.

Mathematical modeling of disease spread and outbreaks has a long and rich history; see
\cite{martcheva2015introduction, frauenthal2012mathematical,pfeiffer2008spatial, nelson2014infectious},
to name just a few. One of the most common classes of these models consists of coupled ordinary
differential equations (ODEs), whose state variables include populations of the host (here, humans)
and the disease carrier, or vector. These populations are further specified as either
\textbf{S}usceptible, \textbf{E}xposed, \textbf{I}nfected, or \textbf{R}ecovered, leading to
thus-named SEIR models.\footnote{Commonly, the model name will specify which sub-populations are
included for both species. For example, an SEIR-SEI model includes host sub-populations of the
susceptible, exposed, infected, and recovered, and vector sub-populations of susceptible, exposed,
and infected.} These models are relatively simple to implement and understand. Model parameters
allow for the specification of transmission rates, incubation times, etc. 

In particular, we investigate the model discrepancy of a well-studied SEIR-SEI model of the Zika
outbreak in Brazil in 2016 \cite{dantas2018calibration}. In previous work, after 
calibration of model parameters, the reduced model captured major tendencies of the outbreak. This
was a major improvement compared to the reduced model with parameter values as suggested by current
literature, which bore almost no usable resemblance to the real epidemic data. However, the
calibrated model was still insufficient to precisely capture the dynamical behavior of the Zika
outbreak.  The current work extends previous works of embedded model discrepancy, used in the contexts of
combustion \cite{morrison2018representing} and ecological models \cite{morrison2019embedded}, to the
current domain of epidemiology. The discrepancy model is embedded within the coupled model
differential equations, and the introduced discrepancy parameters are calibrated to available data.
The enriched model is shown to greatly outperform the original model.

To differentiate the current article from\,\, \cite{morrison2019embedded}, note that that study was
primarily a numerical study over a constrained set of scenarios. The interaction matrices
(determining reduced and true model coefficients) follow a number of assumptions, such as
negative-definiteness, yielding highly well-behaved models. In addition, the actual discrepancy was
known exactly between each reduced model and the corresponding data-generating model; some of that
information was used to further constrain the discrepancy model parameters. Thus, the previous paper
provided no guarantee that this method would work in a highly applied, real-world model scenario
without those strong assumptions.

Although the current modeling scheme only describes a single outbreak \footnote{Note that no
outbreak of Zika was reported in the years following 2016, and thus there is no useful data.}, and
thus is not suitable to describe multiple incidences of the disease, this type of model is useful
for guiding decision and policy makers. For example,\,\, \cite{marnore2016} describes common
questions faced by decision makers, such as how many total people will be infected, or even how to
slow or prevent the outbreak from occurring.  The objective of the present article is to reproduce
with reasonable precision the data of a real epidemic. An appropriate enrichment and calibration of
the model can then yield useful predictions about the dynamical system.

A final complication of this field is that new disease cases are often not reported, causing the
outbreak numbers to appear artificially low. A study from 2018 estimates that as much as 90\% of the
cases are not reported \cite{bastos2018estimating}.  However, as we will see shortly, the issue of
faulty data is insufficient to account for discrepancy between the original model and observations.
At the same time, the issue of under-reported cases does obviously play an important role during the model
validation process.  We try to disentangle the two problems---observational error and model
error---by first considering only model error, and later allowing for both model error and
significant under-reporting. We consider how the enriched model performs in different possible
under-reporting scenarios, such as 10 or 50\% percent.

The rest of the paper is organized as follows. Section \ref{sec:zika} describes the specific model
of the 2016 Brazilian Zika outbreak, and reconstructs previous results as a reference.
Section~\ref{sec:edo} presents the formulation and calibration of the embedded discrepancy operator,
and also corresponding numerical results. Section~\ref{sec:rep} explores the issue of
under-reporting and how well the enriched model might perform given some sample under-reporting
scenarios. A brief concluding discussion is given in Section~\ref{sec:con}.

\section{Zika disease modeling}\label{sec:zika} As mentioned in \S~\ref{sec:int}, a typical approach
to model the spread of infectious disease is with a set of coupled ordinary differential equations.
Here we consider the well-known $SEIR-SEI$ model, which describes coupled growth rates of species of
interest, namely, susceptible, exposed, infected, and recovered humans, as well as susceptible,
exposed, and infected vector.  In the case of Zika (also, Dengue and Yellow fever) in Brazil, this
vector is the \textit{Aedes aegypti} mosquito.

In this section and the next, we assume that the data represents the actual truth. That is, the
modeling objective is to achieve a model consistent with the given data.

\subsection{Model specification}
We follow the SEIR-SEI model discussed by Dantas, Tosin, and Cunha
\cite{dantas2018calibration}, which included species $S_h, E_h, I_h, R_h, S_v, E_v, I_v$. Subscripts
$h$ and $v$ indicate human and vector, respectively. The model also includes a state variable
$C(t)$, which counts cumulative new cases over time. Then, the eight coupled equations are:
\begin{subequations}
\begin{align}
   \frac{dS_h}{dt} &= -\beta_h S_h I_v /N_v\label{eq:z1}\\
   \frac{dE_h}{dt} &= \beta_h S_h I_v /N_v - \alpha_h E_h\\
   \frac{dI_h}{dt} &= \alpha_h E_h - \gamma I_h\\
   \frac{dR_h}{dt} &= \gamma I_h\\
   \frac{dS_v}{dt} &= \delta N_v - \beta_v S_v I_h/N_h - \delta S_v\\
   \frac{dE_v}{dt} &= \beta_v S_v I_h/N_h - (\alpha_v + \delta) E_v\\
   \frac{dI_v}{dt} &= \alpha_v E_v -  \delta I_v\\
   \frac{dC}{dt} &= \alpha_h E_h\label{eq:z8},
\end{align}\label{eq:z}
\end{subequations}
where $N_h$ represents Brazil's human population and $N_v$ represents the vector population.
Nominal values of the interaction rates were determined by a careful literature study. These rates are:
\begin{subequations}
\begin{align}
    \text{Extrinsic incubation period:} \quad \frac{1}{\alpha_v} &= 9.1\\
    \text{Intrinsic incubation period:} \quad  \frac{1}{\alpha_h} &= 5.9\\
    \text{Human infectious period:} \quad \frac{1}{\gamma} &= 7.9\\
    \text{Vector lifespan:} \quad \frac{1}{\delta} &= 11\\
    \text{Mosquito to human infection time:} \quad \frac{1}{\beta_h} &= 11.3\\
    \text{Human to mosquito infection time:} \quad \frac{1}{\beta_v} &= 8.6.
\end{align}
\end{subequations}
Let us collect these model parameters into the vector $\theta$, and let $\theta_n$ refer to the
nominal values given above.

To fully specify this model, it remains to provide initial conditions. These are:\footnote{Although
it might appear above as though some initial conditions are defined in terms of undefined
quantities, this is just an ordering issue. The initial conditions are properly specified in the
following order: $C$, $R_h$, $E_h$, $S_h$, $I_v$, $E_v$, $S_v$.}
\begin{subequations}
\begin{align}
    S_h(0) &= N_h - E_h(0) - I_h(0) - R_h(0) \label{eq:ic1}\\
    E_h(0)&= I_h(0)\\
    I_h(0)&= C(0)\\
    R_h(0)&= 29,639\\
    S_v(0)&= N_v - E_v(0) - I_v(0)\\
    E_v(0)&= I_v(0)\\
    I_v(0)&= 2.2 \times 10^{-4}\\
    C(0) &= 8,201\label{eq:ic8}.
\end{align}
\end{subequations}
Finally, let us call the above model $\mathcal{Z}$ and the state vector $x$, where $x$ is ordered in
the same way as equations~\ref{eq:z1}-\ref{eq:z8} ($x_1 = S_h$, $x_2= E_h$, and so on). Then we may
refer to the above model as:
\begin{equation} \frac{dx}{dt} = \mathcal{Z}(x; \theta). \end{equation} We may also refer to this
    model as the original model, or \emph{reduced} model.

\subsection{Previous results compared to data}\label{ssec:prev} The work in\,\,
\cite{dantas2018calibration} presents a detailed approach to calibrate this model to data. The data
is made available by the Brazilian Ministry of Health \cite{SVS2017} and is available as
supplementary material in\,\, \cite{dantas2018calibration}.  Each data point $d_i, i=1,\dots,52$, gives
the recorded cumulative number of Zika cases at epidemiological week $i$ of the year 2016.  In this section, we
re-plot the results from that paper to serve as an immediate reference and comparison.

\begin{figure}
\includegraphics[width=.4\textwidth]{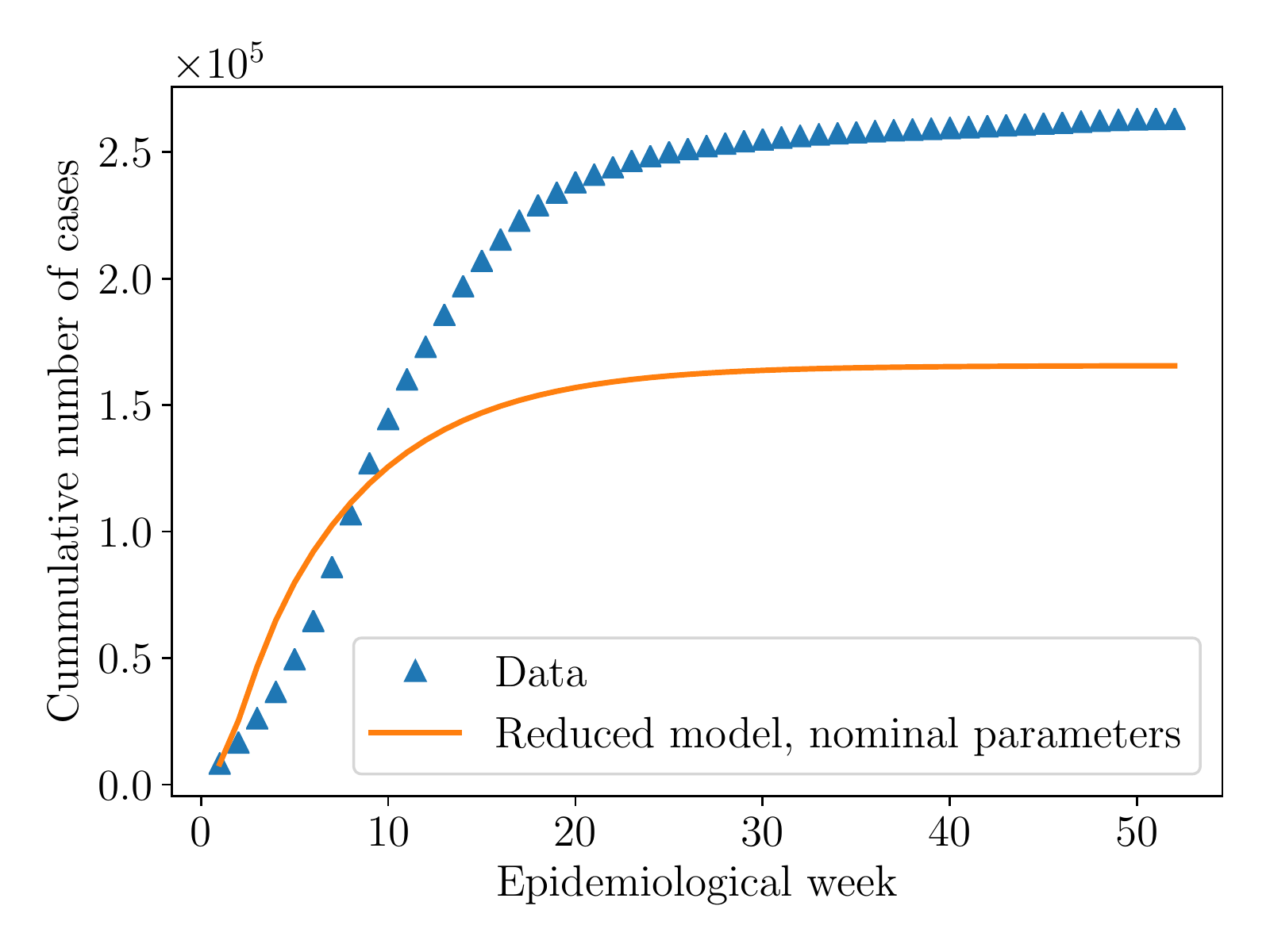}
\caption{\label{fig:nom} Outbreak data and reduced model response using nominal parameter values.}
\end{figure}

First, Figure~\ref{fig:nom} compares the model output to data, using the above model with nominal parameters.
The model with nominal parameter values, $\mathcal{Z}(x; \theta = \theta_n)$, severely underestimates the
outbreak. Note that under-reporting cannot explain the observed discrepancy: higher
reporting rates would only increase this discrepancy.

\begin{figure}
\includegraphics[width=.4\textwidth]{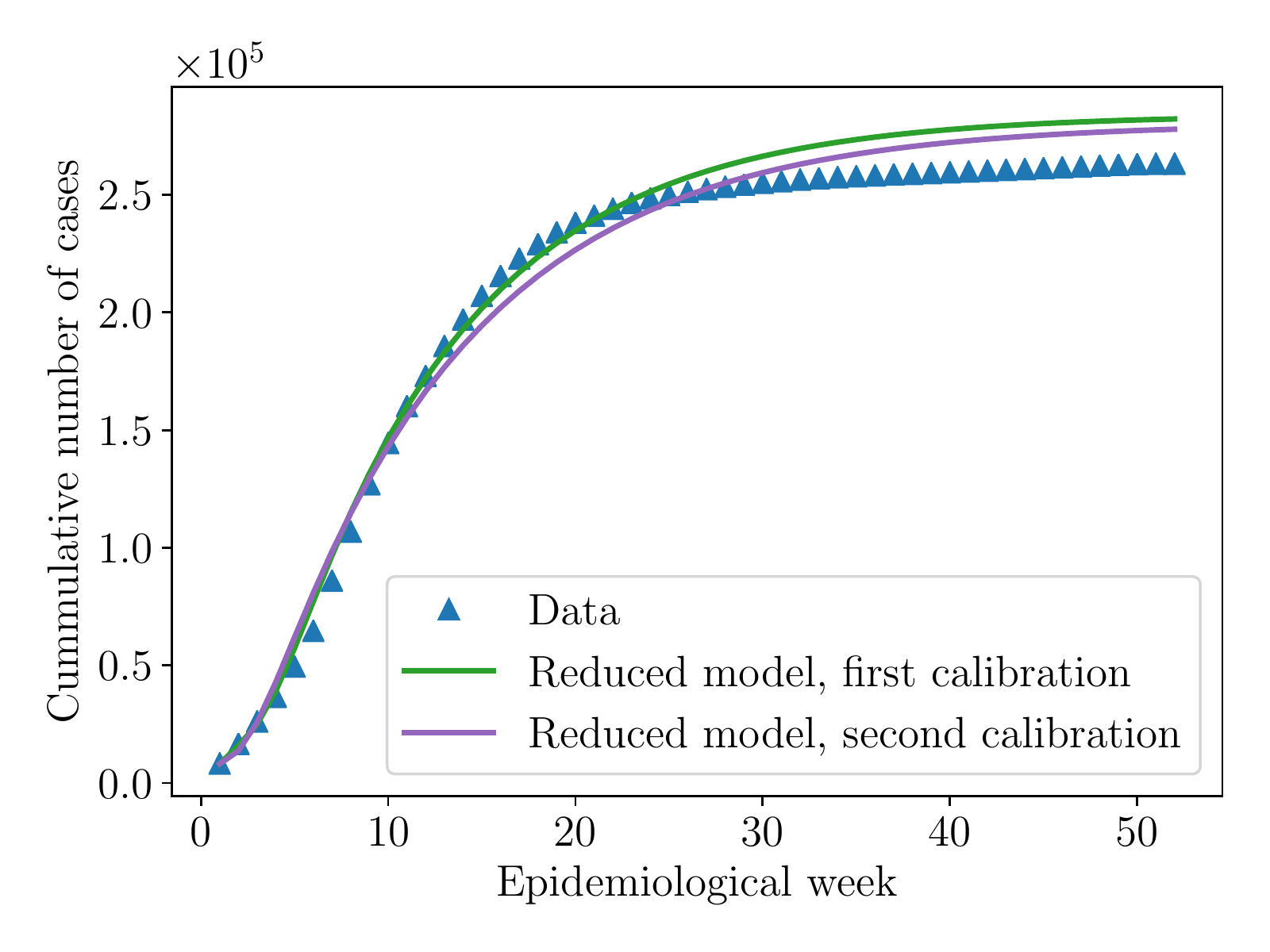}
\caption{\label{fig:cal} Outbreak data and reduced model response using TRR calibrated parameter values.}
\end{figure}
Clearly the reduced model, given $\theta_n$ parameter values, is a poor representation of reality. After
observing such a discrepancy, the authors of\,\,\cite{dantas2018calibration} performed a
sophisticated calibration of the model parameters $\theta$, using the Trust-Region-Reflective (TRR)
method \cite{coleman1996interior, conn2000trust}. Following that method, and using the public data
to calibrate,
two slightly different results are obtained by imposing a different set of constraints on possible
parameters values. The model outputs after both calibrations are shown in Figure~\ref{fig:cal}.
Although the model output is now much closer to the data, still, a detectable inconsistency
persists. To be specific, note that after about week 30, the difference is tens of thousands of new
cases of humans infected by Zika.

From a modeling perspective, the salient point is that the model is unable to capture the dynamical
behavior of the outbreak, even after calibration. Assuming (as we are for now), that the given data
is correct, this suggests that the problem lies with model itself. Indeed, there are several
possible sources of model error, that may impact not only this specific Zika model but many other
epidemiological models. First, these SEIR models are not built from first principles, but rather
from assumptions about interactive behavior, empirical information, and domain scientists' intuition
and experience.  Second, these models provide a continuous, deterministic description of discrete
interactions, which naturally involve some stochasticity \cite{forgoston2009accurate}.  With large
enough populations, though, this should not be a problem. Third, diseases do not spread in a closed
system of host and vector. Rather, the spread of a disease involves other species such as livestock
and non-human primates \cite{childs2019mosquito}. Fourth, other modes of transmission are possible,
such as sexual interaction \cite{coelho2016higher, petersen2016update} and blood transfusion
\cite{motta2016evidence}.  Finally, there could certainly be additional time- or spatially-dependent
effects, such as migrations and local dynamics \cite{chang2020cross,gong2018epidemic}, collective
behavior and time-delayed synchronization dynamics \cite{sun2017behavioral}. Some modelers assume
power-law dynamics of the networks \cite{silva2018activation}, while others use fractional
derivatives to describe relevant dynamics \cite{ghanbari2019analysis}. Instead, this model assumes
time-independent parameters, and only models populations over time, not space. In summary, the
spread of a contagious disease is an incredibly complex problem, and it remains unclear what is
critically missing from the model or how to best improve it directly from epidemiological
information. Here, then, we can turn to the field of model discrepancy to help.

\section{Embedded discrepancy operator}\label{sec:edo}
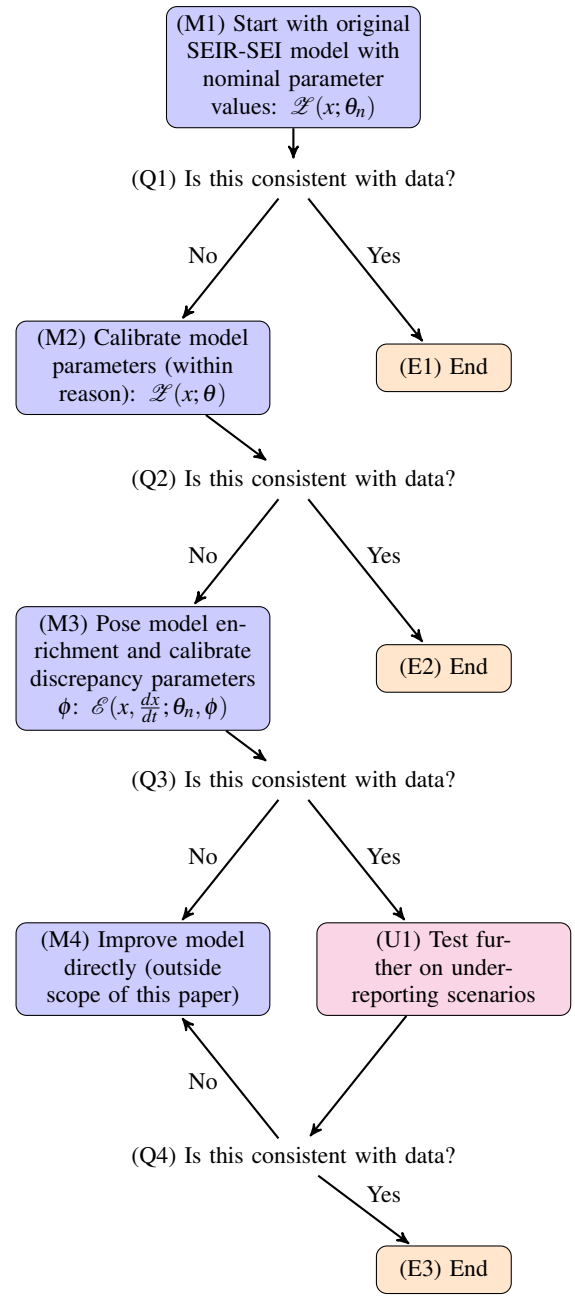
\begin{figure}
    \centering
    %\documentclass[tikz]{standalone}
%\usetikzlibrary{shapes,arrows}
%\begin{document}

\tikzstyle{orangeblock} = [rectangle, draw, fill=orange!20, text width=5em, text
centered, rounded corners, minimum height=2em]%
\tikzstyle{blueblock} = [rectangle, draw, fill=blue!20, text width=10em, text
centered, rounded corners, minimum height=3.5em]%
\tikzstyle{magentablock} = [rectangle, draw, fill=magenta!20, text width=10em, text
centered, rounded corners, minimum height=3.5em]%
\tikzstyle{clearblock} = [rectangle, text width=15em,text centered,minimum
height=1em]%
\tikzstyle{clearblocksmall} = [rectangle, text width=6em,text centered,minimum
height=3em]%
\tikzstyle{line} = [->, >=stealth',shorten >=1pt, thick]%
\tikzstyle{line2} = [<->, >=stealth',shorten >=1pt, thick]%
\begin{tikzpicture}[font=\small]%
%place nodes
    \node at (0,0) [blueblock] (nom) {(M1) Start with original SEIR-SEI model with nominal parameter
    values: $\mathcal{Z}(x; \theta_n)$};
    \node at (0,-1.5) [clearblock] (q1) {(Q1) Is this consistent with data?};
    \node at (2,-4) [orangeblock] (e1) {(E1) End};
    \node at (-2,-4) [blueblock] (cal) {(M2) Calibrate model parameters (within reason):
       $\mathcal{Z}(x; \theta)$};
    \node at (0,-5.5) [clearblock] (q2) {(Q2) Is this consistent with data?};
    \node at (2,-8) [orangeblock] (e2) {(E2) End};
    \node at (-2,-8) [blueblock] (enr) {(M3) Pose model enrichment and calibrate discrepancy parameters
    $\phi$: 
       $\mathcal{E}(x, \frac{dx}{dt}; \theta_n, \phi)$};
    \node at (0,-9.5) [clearblock] (q3) {(Q3) Is this consistent with data?};
    \node at (2,-12) [magentablock] (u1) {(U1) Test further on under-reporting scenarios};
    %\node at (2,-12) [orangeblock] (e3) {(E3) End};
    \node at (-2,-12) [blueblock] (redo) {(M4) Improve model directly (outside scope of this paper)};
    \node at (0,-14.5) [clearblock] (q4) {(Q4) Is this consistent with data?};
    \node at (2,-16) [orangeblock] (e3) {(E3) End};
%draw arrows
\draw [line] (nom) -- (q1);
\draw [line] (q1) -- (cal);
\draw [line] (q1) -- (e1);
\draw [line] (cal) -- (q2);
\draw [line] (q2) -- (enr);
\draw [line] (q2) -- (e2);
\draw [line] (enr) -- (q3);
\draw [line] (q3) -- (redo);
\draw [line] (q3) -- (u1);
\draw [line] (u1) -- (q4);
\draw [line] (q4) -- (e3);
\draw [line] (q4) -- (redo);
%place clear nodes above for load and store
\node at (0,-5.5) [clearblock] {};
\node at (-1.2,-2.5) [clearblock] (n1) {No};
\node at (1.2,-2.5) [clearblock] (y1) {Yes};
\node at (-1.2,-6.5) [clearblock] (n1) {No};
\node at (1.2,-6.5) [clearblock] (y1) {Yes};
\node at (-1.2,-10.5) [clearblock] (n1) {No};
\node at (1.2,-10.5) [clearblock] (y1) {Yes};
\node at (-1.2,-13.5) [clearblock] (n1) {No};
\node at (1.2,-15.0) [clearblock] (y1) {Yes};
%\node at (-1.5,2.5) [clearblock] {$\frac{d\bm{u}^F}{dt}
%\stackrel{?}{\approx}\frac{d\bm{u}^S}{dt}$};

%place nodes below for latencies

%\node at (14,0.5) [clearblocksmall] (memlat) {};
\end{tikzpicture}%

%\end{document}
    \caption{A schematic diagram of the different models and their relationships considered in this
    paper.\label{fig:flow}}
\end{figure}

Before describing the embedded discrepancy operator, we illustrate the overall relationship between
the different models considered in this paper. A schematic diagram is shown in
Figure~\ref{fig:flow}.
As seen in the previous section, after calibrating the model parameters to data, there is still a
significant discrepancy between the model output and the data. That is, the answer to Q2 (and Q1) in
Figure~\ref{fig:flow} is ``No,'' and so we move to state (M3): we model this
discrepancy with the goal of reaching states (U1) and ultimately (E3).

\subsection{Proposed approach: Embedded discrepancy operator}
Previous work has shown that missing dynamics on the right-hand side (RHS) of differential equations
can be approximated with ``extra'' information about the existing state variables
\cite{morrison2019exact, givon2004extracting, hernandez2019algebraic},
such as memory or derivative information. Exploiting this, we pose the following enriched model:
\begin{equation} \frac{dx}{dt} = \mathcal{Z}(x; \theta) + \Delta\left(x, \frac{dx}{dt},
\phi\right) \end{equation}
where
\begin{subequations}
    \begin{align}
        \Delta_{Sh} &= \kappa_1 S_h +  \lambda_1 \frac{dS_h}{dt}\\
        \Delta_{Eh} &= \kappa_2 E_h +  \lambda_2 \frac{dE_h}{dt}\\
        \Delta_{Ih} &= \kappa_3 I_h +  \lambda_3 \frac{dI_h}{dt}\\
        \Delta_{Rh} &= \kappa_4 R_h +  \lambda_4 \frac{dR_h}{dt}\\
        \Delta_{Sv} &= \kappa_5 S_v +  \lambda_5 \frac{dS_v}{dt}\\
        \Delta_{Ev} &= \kappa_6 E_v +  \lambda_6 \frac{dE_v}{dt}\\
        \Delta_{Iv} &= \kappa_7 I_v +  \lambda_7 \frac{dI_v}{dt}\\
        \Delta_{C} &= 0,
    \end{align}
\end{subequations}
with  $\kappa = (\kappa_1, \dots, \kappa_7)$ and $\lambda = (\lambda_1, \dots, \lambda_7)$. That is,
the differential equation for $x_i, i = 1,\dots,7$ in the reduced model is modified by two
additional terms, one linear in $x_i$ and the other linear in $dx_i/dt$. The discrepancy parameters
are collected into the vector $\phi$:
\begin{equation}\phi = (\kappa, \lambda) = (\kappa_1, \dots, \kappa_7, \lambda_1, \dots,
\lambda_7).\end{equation} Note, the RHS for $dC(t)/dt$ is not modified because this function simply
counts the exposed cases as given by the model. A change here would be analogous to modifying model
output itself, not interpretable, and not reliable for any type of decision or prediction.

As mentioned in Section~\ref{sec:int}, this type of discrepancy model can be constrained to
available information about the system. For example, the discrepancy operator
for a combustion reaction in\,\,\cite{morrison2018representing} is constrained to satisfy
conservation of atoms and conservation of energy. In this scenario we do not have such strict
constraints; see\,\,\cite{morrison2019embedded} for constrained operators in similar Lotka-Volterra
models.

All together, the enriched model is 
\begin{subequations}
\begin{align}
    \frac{dS_h}{dt} &= -\beta_h S_h I_v /N_v                           + \Delta_{Sh}\\
    \frac{dE_h}{dt} &= \beta_h S_h I_v /N_v - \alpha_h E_h             +  \Delta_{Eh}\\
   \frac{dI_h}{dt} &= \alpha_h E_h - \gamma I_h                       + \Delta_{Ih} \\
   \frac{dR_h}{dt} &= \gamma I_h                                      +  \Delta_{Rh}\\
   \frac{dS_v}{dt} &= \delta N_v - \beta_v S_v I_h/N_h - \delta S_v   +  \Delta_{Sv}\\
   \frac{dE_v}{dt} &= \beta_v S_v I_h/N_h - (\alpha_v + \delta) E_v   +  \Delta_{Ev}\\
   \frac{dI_v}{dt} &= \alpha_v E_v -  \delta I_v                      +  \Delta_{Iv}\\
   \frac{dC}{dt} &= \alpha_h E_h.
\end{align}
\end{subequations}
Denote the enriched model as $\mathcal{E}(x, dx/dt; \theta, \phi)$, i.e.,
\begin{align}
    \frac{dx}{dt} &= \mathcal{E}\left(x, \frac{dx}{dt}; \theta, \phi\right)\\
    & = \mathcal{Z}\left(x; \theta\right) + \Delta\left(x, \frac{dx}{dt}; \phi\right).
\end{align}
We set $\theta = \theta_n$ and use the same initial conditions as in
equations~\ref{eq:ic1}-\ref{eq:ic8}. The final step to fully specify the enriched model is to
calibrate the discrepancy parameters $\phi$; this step is explained in the following subsection.

\subsection{Calibration details} In contrast with the calibration process
of\,\,\cite{dantas2018calibration} described in \S~\ref{ssec:prev}, here we use a Bayesian
framework\cite{howson2006scientific, jaynes2003probability} to calibrate the discrepancy parameters
$\phi$ given the data $d$.  This allows for the representation of uncertainty about these
parameters, and also how this uncertainty propagates to model output.

Recall that the observations are cumulative cases at each epidemiological week, $d = \{d_i\}, i=1,\dots,52$. For each
$d_i$, let $y_i$ be the corresponding model output. We assume that the measurements are independent
and that the measurement error is additive and Gaussian as such:
\begin{equation}
d_i = y_i + \epsilon, \quad \epsilon \sim \mathcal{N}(0, \sigma_\epsilon^2)\label{eq:eps} \end{equation}
        with standard deviation $\sigma_\epsilon = 5\times10^3$. This standard deviation value seems
        reasonable as the uncertainty in reported values is high, and because the observations are
        on the order of tens to hundreds of thousands.

In the Bayesian framework, the conditional probability density of $\phi$ given the data $d$,
$p_{\text{po}}(\phi|d)$, is called the \emph{posterior} and given as:
\begin{equation}
    p_{\text{po}}(\phi | d) = \frac{p_{\text{li}}(d | \phi)
    p_{\text{pr}}(\phi)}{p_{\text{ev}}(d)}.\label{eq:bayes}
\end{equation}
We specify each term on the RHS above:
\begin{itemize}
    \item \emph{Prior:} The prior density $p_{\text{pr}}(\phi)$ collects the prior knowledge we have about the
        parameters. Specifically, these parameters are assumed independent and uniform in the prior,
        where each $p_{\text{pr}}(\phi_i) = \mathcal{U}(-0.3, 0.15)$, and so
        \begin{equation} p_{\text{pr}}(\phi) = \prod_{i=1}^{14}p_{\text{pr}}(\phi_i) .\end{equation}
    \item \emph{Likelihood:} The likelihood $p_{\text{li}}(d | \phi)$ tells us how likely it is to
        observe $d$, given a particular value of $\phi$. The measurement error model in
        Eq.~\eqref{eq:eps} yields the likelihood function \begin{equation} p_{\text{li}}(d|\theta) =
        \frac{1}{\sqrt{2\pi |\Sigma|}} \exp{\left( -\frac{1}{2}(d-y)^T \Sigma^{-1} (d-y)
\right)},\end{equation} where $\Sigma = \sigma_\epsilon^2 I$.
    \item \emph{Evidence:} The evidence $p_{\text{ev}}(d) = \int
        p_{\text{li}}(d|\phi)p_{\text{pr}}(\phi) d\phi$ gives the probability of
        observing the data $d$. This is typically difficult to compute, but note that it is not a
        function of $\phi$. With a Markov chain Monte Carlo (McMC) approach, the posterior is found
        by computing ratios of the RHS in equation~\ref{eq:bayes} (for different values of $\phi$),
        and so fortunately this term cancels.
\end{itemize}
Under this framework, the discrepancy parameters $\phi$ are calibrated using the \textsc{DRAM}
method, developed by \cite{haario2006dram} and implemented through the library \textsc{QUESO}
\cite{prudencio2012parallel}. The complete code for this project is available here:
\texttt{https://github.com/rebeccaem/zika} \cite{morrison2020zikacode}.

Table~\ref{tab:phi} presents posterior means and standard deviations for each of
the fourteen marginal posterior densities of the discrepancy parameters (as histograms).
\begin{table}
    \caption{\label{tab:phi}Information from the parameter posterior $p_{\text{po}}(\phi | d)$.}
\begin{tabular}{|c|c|c|}
    \toprule
Parameter & Posterior & Posterior\\
    &mean &standard deviation\\
\midrule
$\kappa_1$ & -0.04 & 0.005\\
$\kappa_2$& -0.26 & 0.02\\
$\kappa_3$& 0.10& 0.02\\
$\kappa_4$ & -0.04 & 0.13\\
$\kappa_5$& 0.07 & 0.01\\
$\kappa_6$& 0.11 & 0.02\\
$\kappa_7$& 0.10 & 0.01\\
$\lambda_1$& 0.00 & 0.09\\
$\lambda_2$& -0.15 & 0.08\\
$\lambda_3$& -0.18 & 0.08\\
$\lambda_4$& -0.15 & 0.09\\
$\lambda_5$& -0.02 & 0.10\\
$\lambda_6$& -0.15 & 0.10\\
$\lambda_7$& -0.06 & 0.09\\
    \bottomrule
\end{tabular}
\end{table}

\subsection{Numerical results}\label{sec:res} 
\begin{figure}
\includegraphics[width=.4\textwidth]{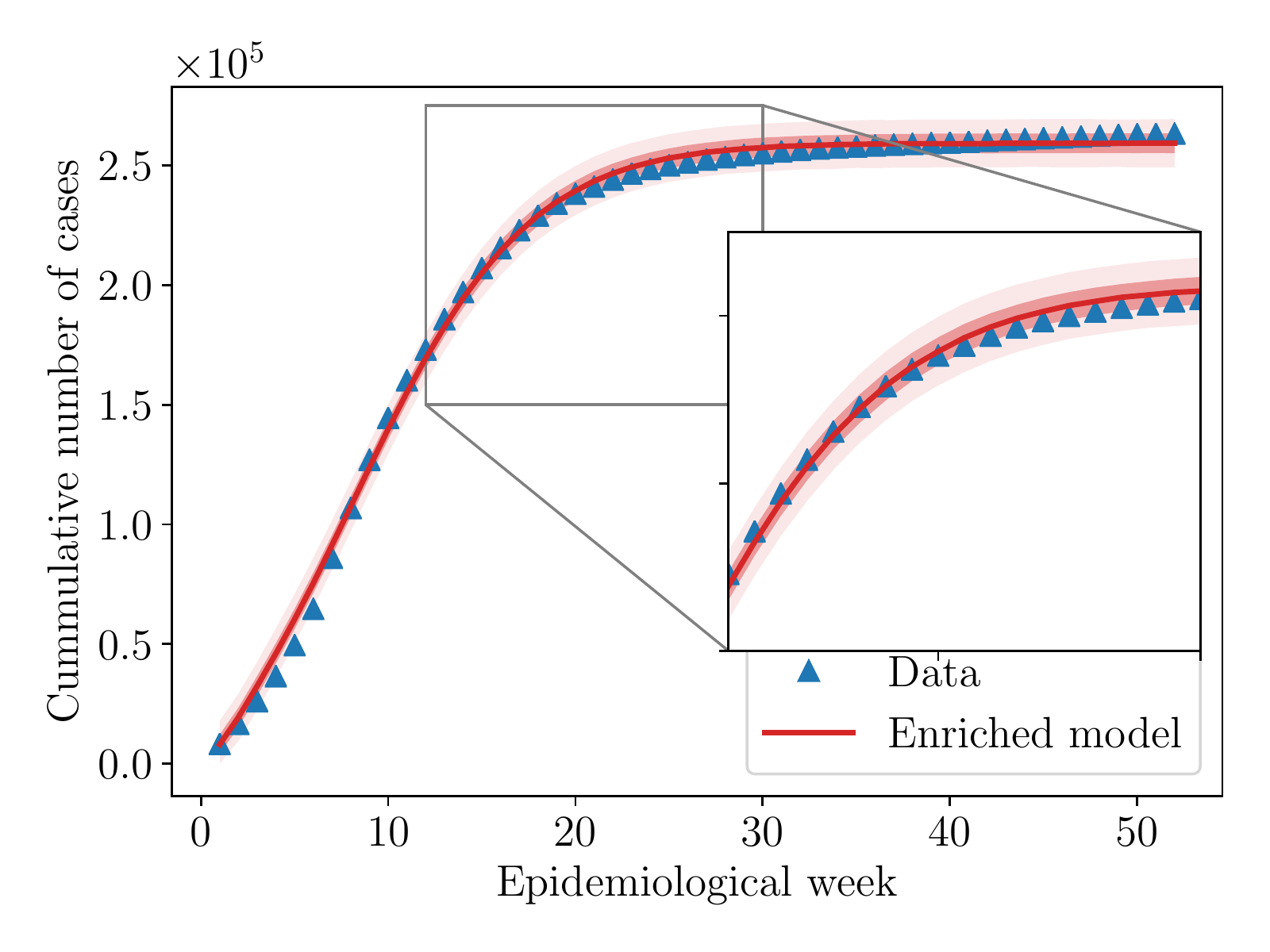}
\caption{\label{fig:enr} Outbreak data and enriched model response.}
\end{figure}

Figure~\ref{fig:enr} shows the enriched model response
compared to the data. Uncertainty in discrepancy parameters $\phi$ is propagated through to model
output: the thick center line shows the median response, the darker band shows the 50\% confidence
interval, and the lighter band the 95\% confidence interval (CI).  Importantly, all observations are
in fact captured by the 95\% CI.

\begin{figure}
  \begin{subfigure}{.4\textwidth}
\includegraphics[width=\textwidth]{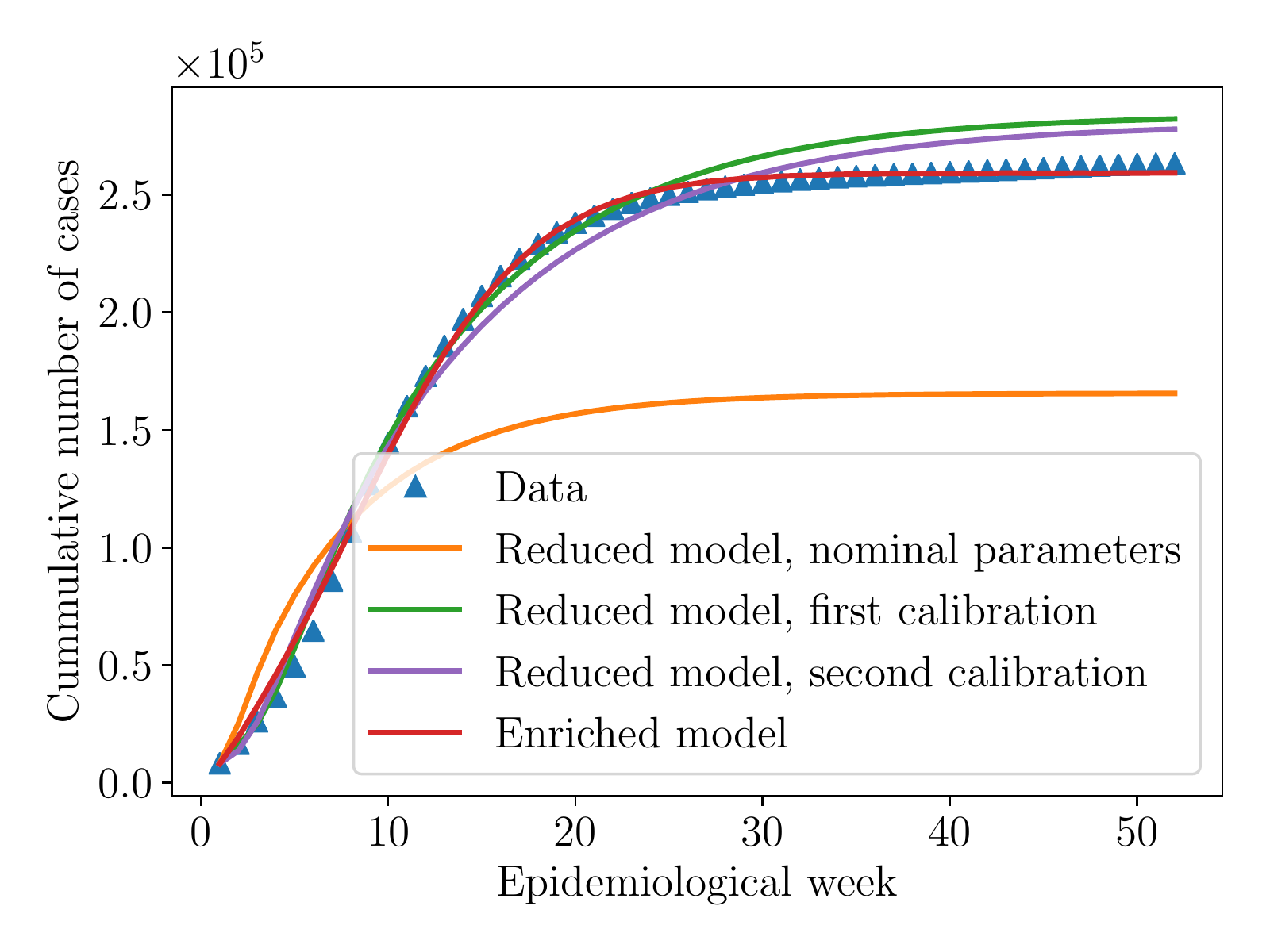}
\caption{\label{fig:all}}
  \end{subfigure}
  \begin{subfigure}{.4\textwidth}
\includegraphics[width=\textwidth]{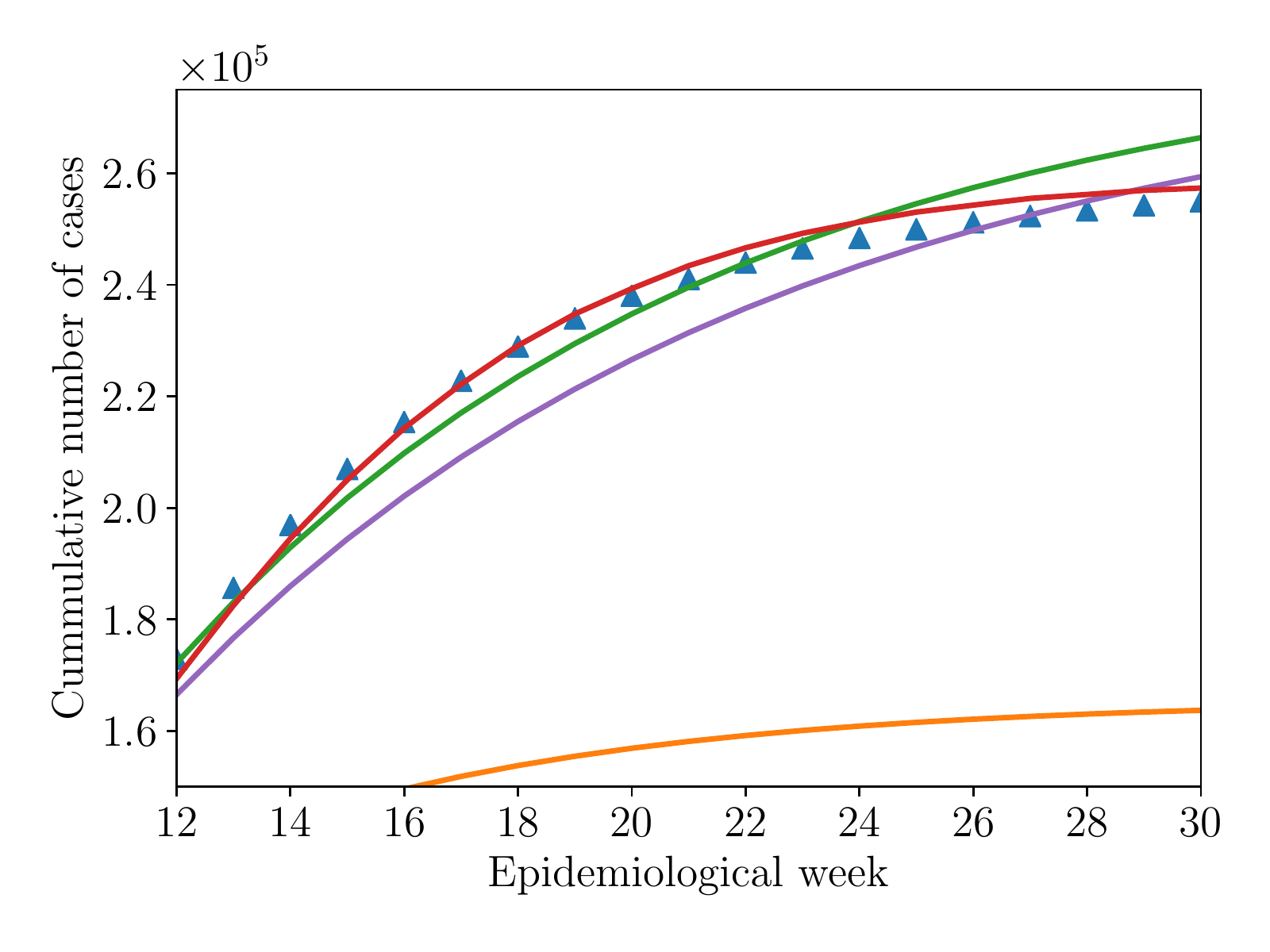}
      \caption{\label{fig:allzoom}}
  \end{subfigure}
    \caption{Outbreak data compared to all  model responses. Figure (b) is zoomed into weeks
    12-30.}
\end{figure}

For comparison's sake, Figure~\ref{fig:all} presents at once all model responses considered in this
paper, and Figure~\ref{fig:allzoom} shows the same, but zoomed into weeks 12-30. (For visualization
purposes, only the median line is shown for the enriched model.) The enriched model is clearly an
improvement.

\subsection{Interpretation}\label{ssec:interp} The embedded discrepancy operator can be interpreted
from two different points of view. The first is a more mathematical lens, although not disconnected
from the physics: we interpret the discrepancy operator as a linear feedback signal. The second, in
contrast, relies on an epidemiological basis: here we interpret the corrections made by the
discrepancy operator as effects due to causes of biological origin.

This second point of view is especially interesting for the goal of elucidating potential
deficiencies in the baseline model. As a wide range of issues must yet be explored to obtain a
consistent epidemiological interpretation, this line will not be addressed in this manuscript, but
will be the topic of future work. Instead, the first perspective is explored further below.

In light of theory of systems with linear feedback, the discrepancy operator is a linear combination
of the system state and its first order time-derivative, thus defining a signal that feeds the
original nonlinear system with information from the present state and its rate of change. Roughly
speaking, the parameters of the enrichment can be seen as ``gains'' that adjust to drive the
epidemic curve generated by the model towards the real observational curve.  These parameters are
identified via Bayesian inference, with prior distributions that admit negative and positive values:
thus the gains can define both negative and positive feedbacks. The realizations of the enriched
dynamical system may then admit a superposition between negative and positive feedbacks, generating
a kind of competition between the model's stimuli signals. This competition stabilizes some
coordinates of the state vector and destabilizes others, while the global effect materializes in the
corrected (enriched) epidemic curve.

To understand more deeply why this competition between corrective signals produces such an effective
correction, consider the injection and removal of information (i.e., energy) in the system, as well
as its flow between the different coordinates of the state (groups of human and mosquito
populations).  To make an analogy with the dynamics of mechanical oscillators, the feedback effects
proportional to the state derivative produce a kind of ``viscous force,'' which introduces (via positive
feedback) or removes (via negative feedback) information into or from the epidemiological state
variables. Furthermore, the terms proportional to the system state correspond to a kind of
``restoring force,'' which redistributes information among the different population groups. The
intensity of this additional information flow between the different coordinates of the system state
is controlled by the new time scales induced by the feedback signals, which are nonlinear functions
of the gains $\kappa_i$ and $\lambda_i$, $i = 1,\dots, 7$.

It should be noted that the approach presented in the paper is not control theory in the literal
sense, since no information from the biological system is obtained in real-time, nor is an action
signal sent to the real system to adjust its epidemic curve trajectory. Therefore, the observability
and controllability issues related to the real system are not taken into account. The notion of
duality between parameters and gains is explored above only as a way to pave an initial reasonable
interpretation of how the discrepancy operator acts to correct the model's response, by promoting
additional information flows between the different compartments of the populations.

%\begin{figure}
%\includegraphics{}%
%\caption{\label{}}%
%\end{figure}

\section{Effects of under-reporting}\label{sec:rep}
\begin{figure}
\includegraphics[width=.4\textwidth]{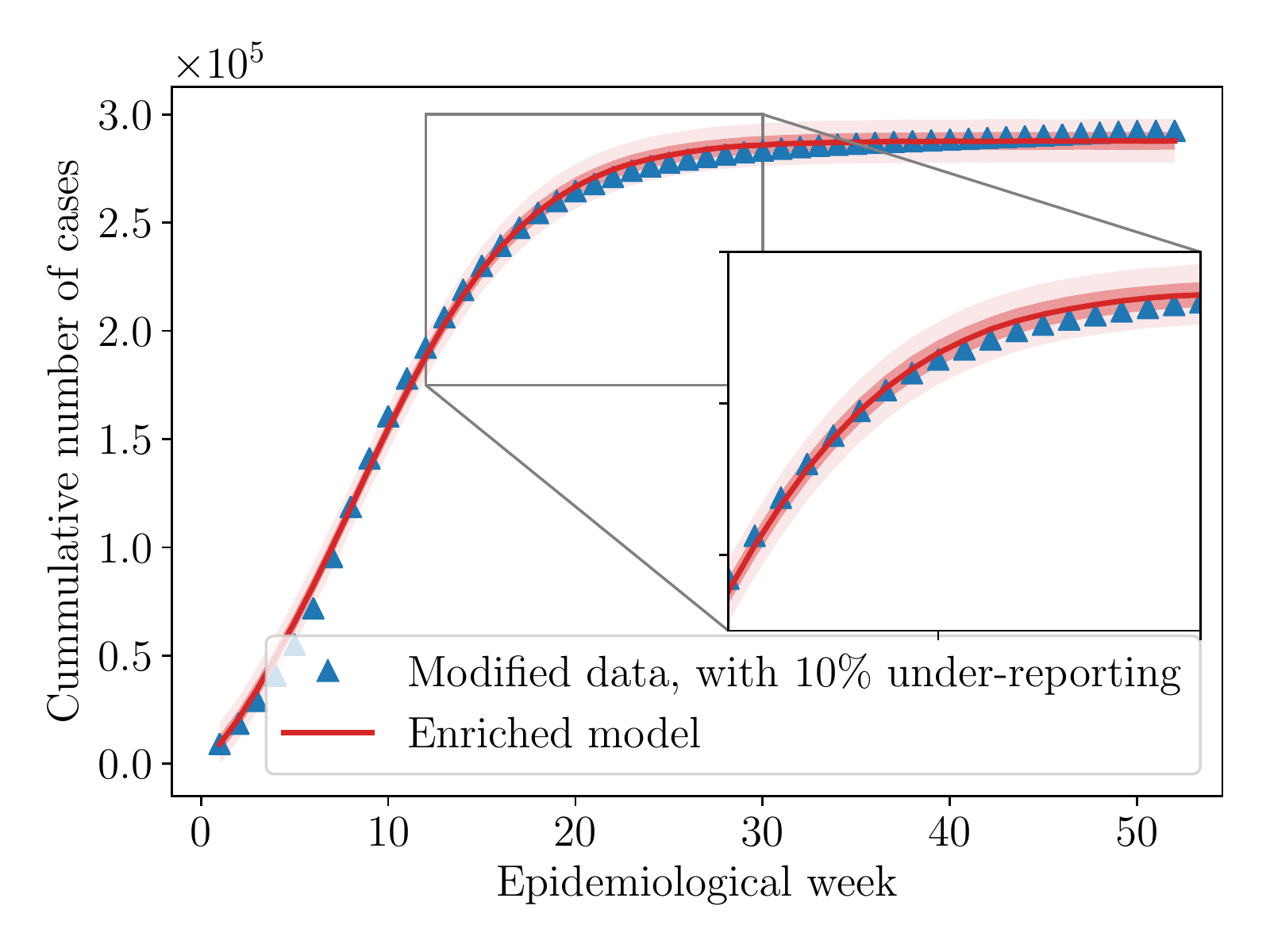}
\caption{\label{fig:rep10p} Modified outbreak data, assuming 10\% under-reporting, and enriched model
    response.}
\end{figure}

Now let us also consider the scenario that the data is in fact under-reported. First, suppose 10\%
of cases are not reported, so that $d_i = .9 d_i^*$, where $d_i^*$ represents the value of
observations we would expect without under-reporting. (This is not claiming $d_i^*$ is the exact
true value, as we still expect unbiased measurement error.) The discrepancy parameters are
re-calibrated, and the corresponding model response is shown in Figure~\ref{fig:rep10p}. Again, all
(modified) observations are captured by the enriched model response.

\begin{figure}
\includegraphics[width=.4\textwidth]{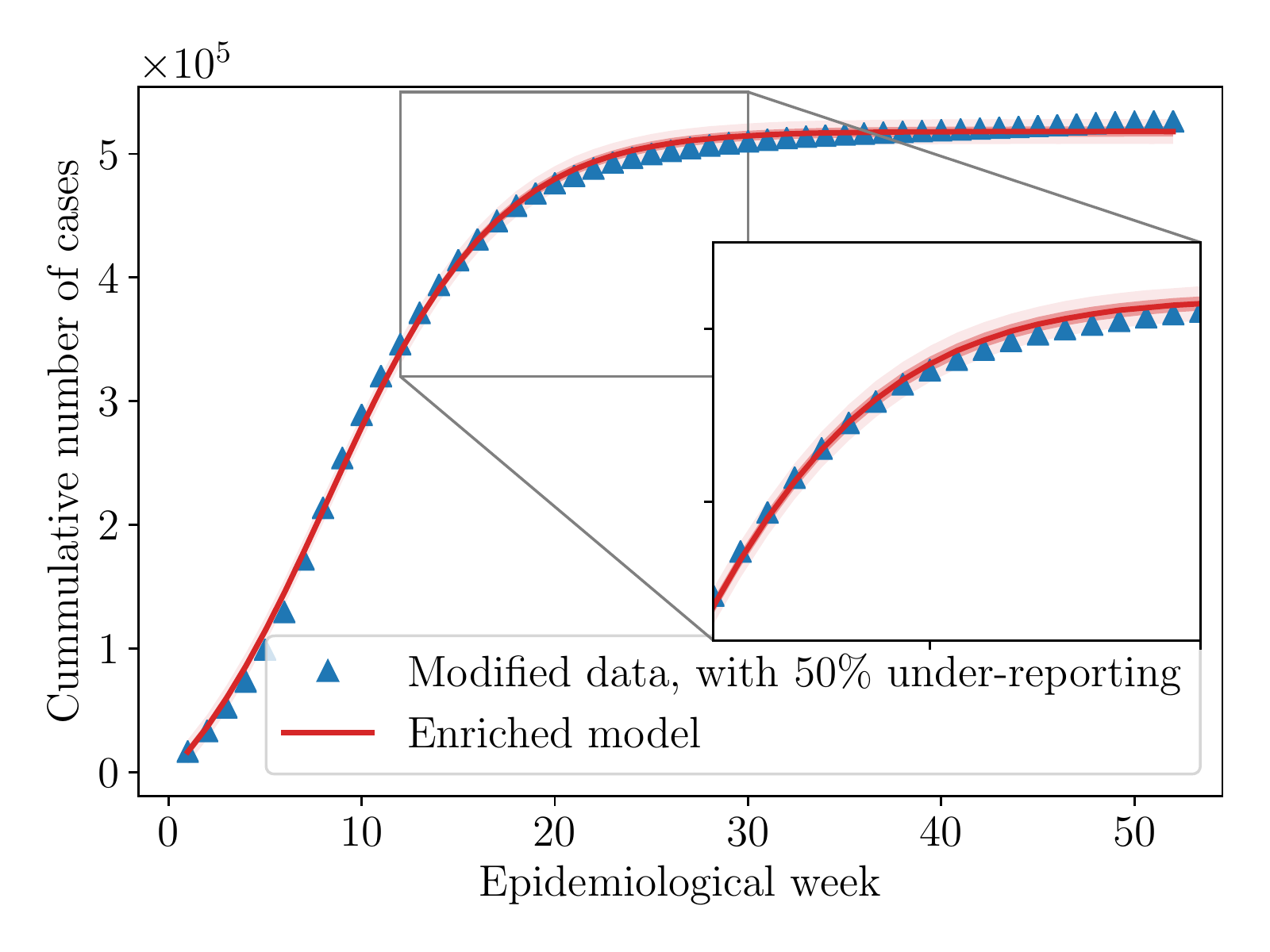}
\caption{\label{fig:rep50p} Modified outbreak data, assuming 50\% under-reporting, and enriched model
    response.}
\end{figure}

Finally, we suppose that only 50\% of cases are reported, so that $d_i = 0.5 d_i^*$. These results
are shown in Figure~\ref{fig:rep50p}. Even here, the enriched model adapts to this scenario and
covers the dynamical behavior of the outbreak in this highly under-reported scenario.

\section{Conclusion}\label{sec:con} This work presents an initial endeavor to represent the model
discrepancy of an epidemiological system, namely, the 2016 Brazilian Zika outbreak. Preliminary
results are promising---compared to the original model, the embedded discrepancy operator greatly
improves the consistency between model output and available observations.

The general applicability of this method to other epidemiological models is best understood in two
parts. One one hand, the formulation of the enriched model equations is immediately applicable to
another model, if that model is also comprised of a set of ODEs. That is, nothing prevents a modeler
from testing the proposed model enrichment framework in that case. On the other hand, the particular
details of the calibration, and whether or not this approach is in fact able to capture the
discrepancy between the original model and the data may depend on domain specific information.
Future studies will test this approach in other outbreak data sets.

Many other open questions remain. In Section~\ref{ssec:interp}, we discuss two possible
interpretations of the embedded discrepancy operator. First, the linear terms added into the
differential equations resemble a type of linear feedback, with one term proportional to the
state variable, and one a differential ``control'' term. In this case, the discrepancy operator is not
actually controlling the real system itself, but driving the model system to the target (epidemic
data). Second, and perhaps more importantly, would be a physiological interpretation of the
discrepancy terms. While beyond the scope of this paper, a deep exploration of interpretability,
connections to linear feedback theory, and an explanation of these discrepancy terms in a
physiological sense will be the subject of immediate future work.

Related to the point above, we would also like to understand what the calibrated discrepancy
operator implies about the missing dynamics of the reduced model. That is, can we use the learned
discrepancy model to infer what the reduced model is most critically lacking? This question is
currently under study, also in the context of ecological models (which have a similar structure, as
sets of coupled ODEs). Doing so would allow the use of these embedded operators to function as a
type of modeling tool, as opposed to only a model correction.

Finally, this study would be perhaps more convincing with more trust-worthy data. How to achieve
this, though, is just as complex a problem as the epidemiological system itself, as it involves
accessibility to healthcare in remote regions, public awareness of mandatory reporting policies, and
incentives and rewards for timely reporting of a communicable disease.

%\subsection{}
%\subsubsection{}

% If in two-column mode, this environment will change to single-column format so that long equations can be displayed. 
% Use only when necessary.
%\begin{widetext}
%$$\mbox{put long equation here}$$
%\end{widetext}

% \begin{figure}
% \includegraphics{}%
% \caption{\label{}}%
% \end{figure}

% Tables may be be put in the text as floats.
% Here is an example of the general form of a table:
% Fill in the caption in the braces of the \caption{} command. Put the label
% that you will use with \ref{} command in the braces of the \label{} command.
% Insert the column specifiers (l, r, c, d, etc.) in the empty braces of the
% \begin{tabular}{} command.
%
% \begin{table}
% \caption{\label{} }
% \begin{tabular}{}
% \end{tabular}
% \end{table}

% If you have acknowledgments, this puts in the proper section head.
\begin{acknowledgments}
    The authors acknowledge Prof. Davi Ant\^{o}nio dos Santos (ITA Brazil) for very helpful discussions,
    especially regarding the linear feedback aspect of this work.
The second author acknowledges the financial support given by the Brazilian agencies
    Coordena\c{c}\~{a}o de Aperfei\c{c}oamento de Pessoal de N\'{\i}vel Superior - Brasil (CAPES) -
    Finance Code 001, and the Carlos Chagas Filho Research Foundation of Rio de Janeiro State
    (FAPERJ) under grants 210.021/2018 and 211.037/2019.
\end{acknowledgments}

\section*{Data availability statement}
 The data are available in a database maintained by the Brazilian Ministry of Health \cite{SVS2017}
    and also available as supplementary material in\,\, \cite{dantas2018calibration}. All data (and
    code) needed to reproduce results in this paper are also included in the code base\,\,
    \cite{morrison2020zikacode} with \texttt{doi:10.5281/ZENODO.3666845}.
% Create the reference section using BibTeX:
\bibliography{references}
%\section*{References}

\end{document}